\documentclass[a4paper,11pt]{article}
\usepackage{jinstpub} 
\usepackage{hyperref}
\usepackage{url}
\usepackage{lineno}

\title{\boldmath Testing the limits of ITkPixV2: the ATLAS inner tracker pixel detector readout chip}

\author[a,b,1]{L. Le Pottier, \note{Corresponding author.}}
\author[b]{T. Heim}
\author[b]{M. Garcia-Sciveres,}
\affiliation[a]{University of California Berkeley, Berkeley, CA 94720}
\affiliation[b]{Lawrence Berkeley National Lab, Berkeley, CA 94720}

\emailAdd{luclepot@lbl.gov}

\abstract{The ITkPixV2 chip is the final production readout chip for the ATLAS Phase 2 Inner Tracker (ITk) upgrade at the upcoming High-Luminosity LHC (HL-LHC). Due to the extraordinarily high peak luminosity at the HL-LHC of $5 \times 10^{34}$ cm$^{-1}$s$^{-1}$, ITkPixV2 must meet significant increases in nearly all design requirements compared to the current ATLAS Inner Detector (ID), including a 10x increase in trigger rate, a 7.5x increase in hit rate, a 3x increase in radiation tolerance, and a 12.5x decrease in pixel current draw per unit area, all while maintaining a similar power per unit area as present pixel detectors. Here we present the first measurements of the ITkPixV2 chip operated at the limits of the full chip design requirements, including in particular a measurement of the activity-induced current of the chip as a function of increasing hit rate.}

\keywords{Tracking detectors; detector front-end electronics; HL-LHC upgrade detector physics}

\begin{document}
\maketitle

\flushbottom

\section{Introduction}\label{sec:intro}

To handle the increased total luminosity and increased number of collisions per beam crossing during the High-Luminosity LHC (HL-LHC), the ATLAS experiment will completely replace its inner detector system. The ATLAS Inner Tracker (ITk) upgrade will consist of both pixel and strip detector systems. The final version of the ITk Pixel detector chip is the ITkPixV2 chip, which has been developed over the course of over a decade by the CERN RD53 collaboration~\cite{rd53_collaboration}, along with the CROC-V2 chip for the CMS experiment, meeting similar requirements.
Here we present the first measurements of ITkPixV2 performance at the limit of the ATLAS ITk's upgrade requirements.

\section{The ITkPixV2 chip}\label{sec:chip}

ITkPixV2 is a 65nm feature size CMOS technology chip, the full description of which is provided in~\cite{rd53c_manual}.
It uses the RD53C framework, a generic pixel readout chip framework that can be used to build chips with different sizes and the same digital and analog pixel blocks. 

\begin{table}[!b]
\centering
\caption{Summary table of selected ITkPixV2 final design requirements from~\cite{rd53b_design_reqs}, compared with the currently operating ATLAS Insertable Barrel Layer (IBL) readout chip FE-I4, from~\cite{abbott_production_2018}.\label{tab:reqs}}
\smallskip
\begin{tabular}{c|c|c|c}
    \hline
    Constraint & FE-I4 Chip & ITkPixV2 & Change \\
    \hline
    Chip Size & 2x2 cm$^2$ & 2x2 cm$^2$ &  \\
    Pixel Size & 50x250 $\mu$m$^2$ & 50x50 $\mu$m$^2$ & \textbf{5x} smaller \\
    Hit Rate & 0.4 GHz/cm$^2$ & 3 GHz/cm$^2$ & \textbf{7.5x} higher \\
    Trigger Rate & 100 kHz & 1 MHz & \textbf{10x} higher\\
    Trigger Latency & 6.4 $\mu$s & 12.8 $\mu$s & \textbf{2x} longer\\
    Radiation Tolerance & 300 MRad & 1 GRad & \textbf{3x} higher\\
    Current Draw & 20 $\mu$A / pixel & 8 $\mu$A/pixel & \textbf{2.5x} lower \\
    \hline
\end{tabular}
\end{table}

The chip consists of a pixel matrix and a chip bottom, and the pixel matrix is built of repeated 8 by 8 pixel cores of size 400 $\mu$m by 400 $\mu$m tiled in rows and columns. ITkPixV2 has 48 core rows and 50 core columns with a 50 $\mu$m by 50 $\mu$m pixel pitch, for a total matrix size of 384 x 400 pixel. The chip uses a differential analog front end described in detail in~\cite{rd53a_implementation}.

The design requirements for ITkPixV2 are a significant departure the previous ATLAS Pixel front-end readout ASIC.
These are summarized from~\cite{rd53b_design_reqs} in table~\ref{tab:reqs}; note in particular the hit rate of 3 GHz/cm$^2$ (7.5x increase), trigger rate of 1 MHz (10x increase), and pixel current draw of 8 $\mu$A/pixel (2.5x decrease).
The ITk Pixel system is also expected to have large differences in the requirement for chip output bandwidth as a function of chip position in the detector
, with the data rate difference between the highest rate (3.24 Gbps) and lowest rate (160 Mbps) chips reaching a factor of 20x~\cite{expected_data_rates}.
To accommodate this, ITkPixV2 has 4 readout lanes per chip with a maximum of 1.28 Gbps per lane, for a total theoretical data rate of 5.12 Gbps per chip.

Realistic testing of any one of ITkPixV2 design requirements should be done while simultaneously meeting the others; for instance, measuring the maximum digital current consumption of the chip means reaching the maximum hit rate, trigger rate, and output data rate at the same time. Here we present a measurement of the digital current increase as a function of increasing hit rate, at the target maximum trigger rate of 1 MHz and trigger latency of 12.5 $\mu$s.

\begin{figure}[!b]
\centering
\includegraphics[width=.9\textwidth]{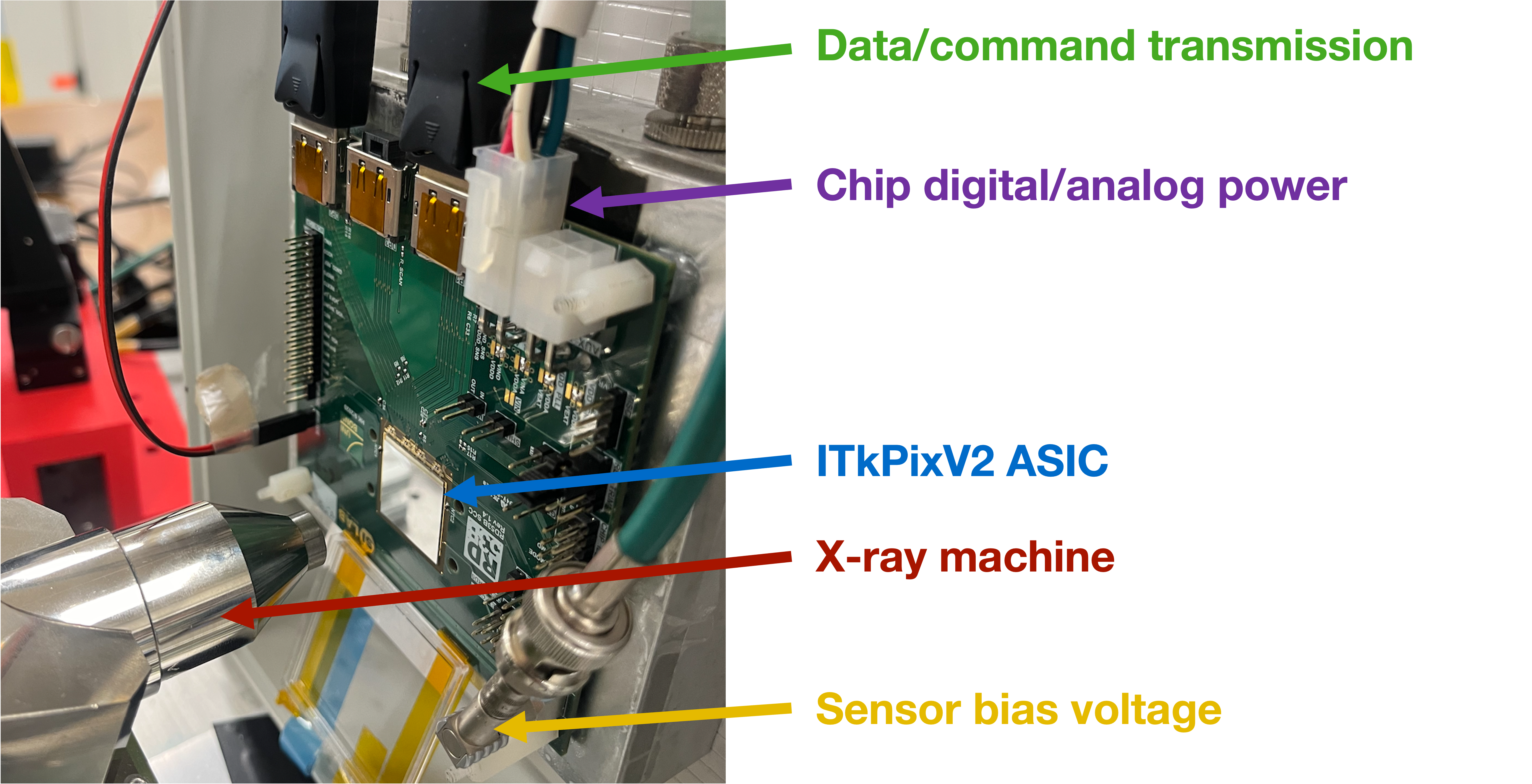}
\caption{X-ray machine and ASIC positioning.\label{fig:setup}}
\end{figure}

\section{Experimental setup}\label{sec:setup}

These tests were performed at the Santa Cruz Institute for Particle Physics (SCIPP), using a single ITkPixV2 chip module on a custom PCB carrier card. Figure~\ref{fig:setup} shows the experimental setup.
To induce high hit rates in the chip we used the Amptek mini-X2 X-ray tube~\cite{mini_x2} with a Silver filament. This X-ray tube produces X-rays with a wide range of energies, and a flux directly proportional to the X-ray machine current.
The X-ray tube was operated at a bias voltage of 40 kV, with current ranging from 0 to 200 $\mu$A. To obtain a relatively uniform X-ray flux across the chip surface, we did not use any collimator on the output of the X-ray tube.

Chip digital and analog power were provided and monitored using a Rhode\&Schwarz HMP4040 power supply, providing an accuracy of $\pm$ ($0.1\% + 2$mA)~\cite{hmp4040}.
The supply voltage was fixed at 1.6V, which is the expected unregulated input value during operation in the serial power chain configuration that will be used during operation, as per~\cite{samy_serial_2024}.

Chip bias voltage was set and monitored with a Keithley 2400 to an accuracy of 0.012\%~\cite{keithly}.
The chip itself carried a 100~$\mu$m thick sensor produced and bump bonded to the ASIC by Micron Technology Inc., with a bias voltage of 35V applied in all tests.
The chip was mounted on a cooling chuck operated at 10~$^\circ$C, and the on-chip NTC temperature reading was approximately 18~$^\circ$C throughout.

\begin{figure}[htbp]
\centering
\includegraphics[width=.45\textwidth]{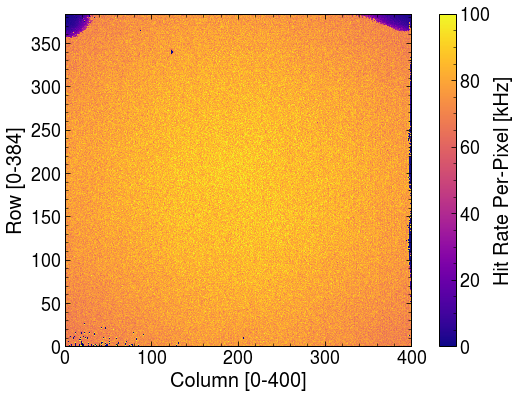}
\qquad
\includegraphics[width=.48\textwidth]{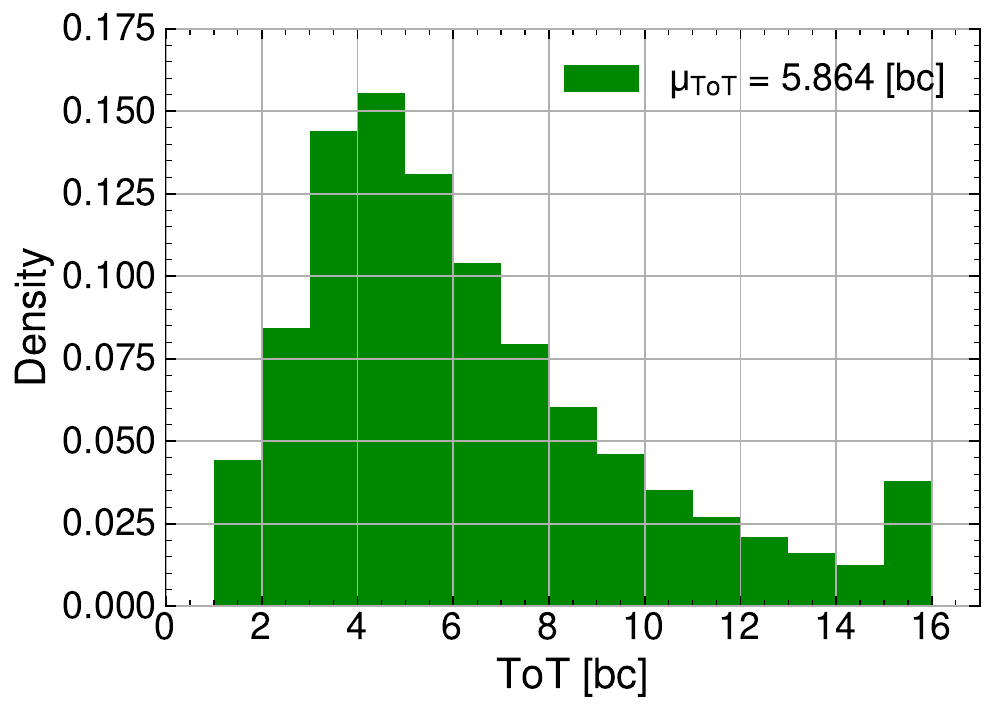}
\caption{Hit rate profile (left) and time over threshold (ToT) distribution (right) for a single trial with our X-ray setup, with tube current $I = 70~\mu$A and an average hitrate of 3.1 GHz/cm$^2$. In the ASIC 1 bc represents a single cycle of the internal chip clock running at 40 MHz, meaning 1~bc 
$\sim$ 25 ns.\label{fig:profile}}
\end{figure}

Figure~\ref{fig:profile} shows the hit rate and time over threshold (ToT) distributions for a single run.
The hit rate is relatively evenly distributed across the chip, with some dead pixels at the corners and edges of the chip due to imperfect bump bonding of the sensor.
We decreased the feedback current for the differential front end until we obtained average ToT of about 6, matching what we would expect from minimum ionizing particles produced in collisions at the LHC.

\begin{figure}[!b]
\centering
\includegraphics[width=.8\textwidth]{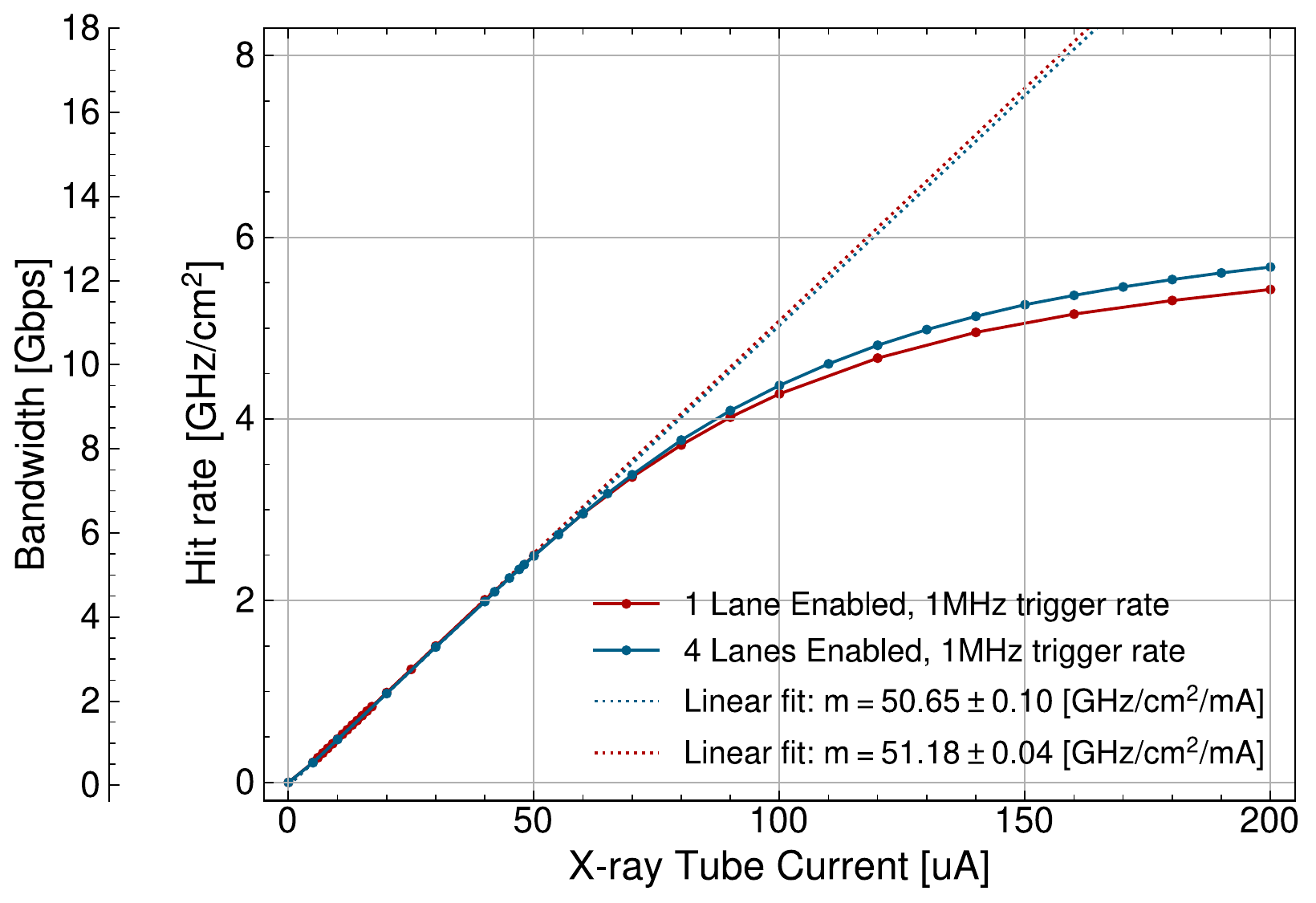}
\caption{Hit/bit rates as a function of X-ray tube current. Linear fits are shown, with slope $m$ defined fitted slope in the region with hit rate $<$ 1 GHz/cm$^2$. These fits are extrapolated to show the relative hit loss at high X-ray tube current values.\label{fig:correction}}
\end{figure}

\section{Results}\label{sec:results}

Measurements of hit rate as a function of tube current were made for the case of both 1 and 4 enabled lanes, corresponding to 1.28 and 5.12 Gbps maximum data transmission rates, respectively. Prior to analyzing the current results, a correction factor must be applied to the data rate and hit rates.


ITkPixV2 was designed to have less than 1\% hit loss due to memory buffer overflows at a hit rate 3 GHz/cm$^2$ with a trigger latency of 12.5 $\mu$s.
As the hit rate starts to exceed the chip design requirements, the buffer overflow hit loss should increase.
To compensate for this effect, fit factors were applied to obtain the true hit and data rates in the chip and are shown in Figure~\ref{fig:correction}.
This factor comes out to about 50.65 GHz/cm$^2$/mA, and can be directly applied as a correction to both the hit rate and the bandwidth at each data point.





\begin{figure}[!b]
\centering
\includegraphics[width=.8\textwidth]{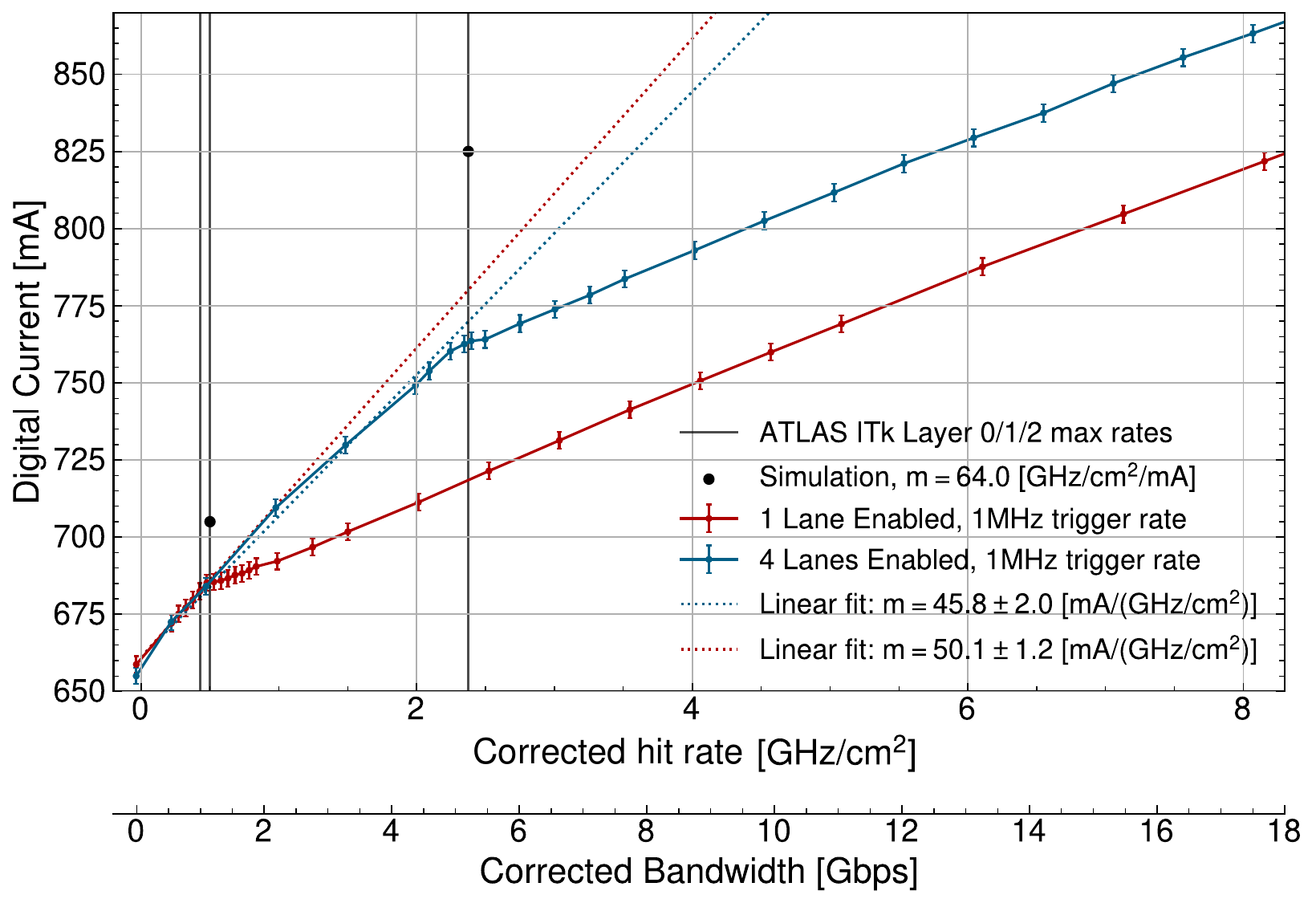}
\caption{Digital current as a function of hit rate. Linear fits are shown with slope $m$ corresponding to the fitted slope in the region before the ``knee:" 2 GHz/cm$^2$ for the 4-lane configuration, and 0.45 GHz/cm$^2$ for the 1-lane configuration.\label{fig:cur}}
\end{figure}

Figure~\ref{fig:cur} shows the final measurement of the digital current as a function of hit rate in the chip, for both the 1-lane (red) and 4-lane (blue) cases. In both cases, we observe a linear increase in the current consumption until we reach the bandwidth limit of 1 Gbps (1-lane) and 4.5 Gbps (4-lane), both within the 80\% safety factor chosen for ITkPixV2.
It is notable that the hit rate at these bandwidth limits is less than the design specs of the chip - 2.45 GHz/cm$^2$ for the 4-lane configuration.
This is in fact the expected behavior: X-rays typically produce clusters with a size of 1, which is the worst case for lossless compression of hit information as is done in RD53. The encoding for single clusters is simulated in~\cite{benchmarking_rd53b} to scale as 1.796 Gbps/(GHz/cm$^2$).
In our X-ray measurements, we measured this empirically to be 1.819 $\pm$ 0.005 Gbps/(GHz/cm$^2$), which matches extremely well.
For ITk, we expect average cluster sizes between 2-4 throughout the detector. Using the encoding scaling given in~\cite{benchmarking_rd53b} for cluster sizes of 2-4, we can estimate our rate limit to be 3.34-3.55 GHz/cm$^2$ during operation -- well within the ITkPixV2 design specs.

After this bandwidth limit is reached, we continue to see an increase in digital current from the pixel matrix  - while hits are no longer able to be read out, their pulses and ToTs are still counted by the pixels on the matrix, increasing digital current draw. 
The slope of the ``pixel matrix'' current increase is identical for both 1 and 4 lane configurations. 

Figure~\ref{fig:cur} also displays fits to the pre-limit regions of the curve with dotted lines, showing that the lane configurations also roughly agree on the slope of the digital current increase prior to the bandwidth limit, at about 45 mA/(GHz/cm$^2$).
This is slightly below the simulated increase of 64 mA/(GHz/cm$^2$), and well within the design specs of ITkPixV2.

\section{Conclusions}

We have presented the first measurements of the ITkPixV2 chip operated at full design specs for hit rate, trigger rate, and latency.
These include measurements of the activity-induced digital current for any iteration of the ITkPix chip, the chip bandwidth limit for both the 1-lane and 4-lane configurations, and the memory efficiency of the chip as a function of hit rate.

\acknowledgments

Thank you to Simone Mazza, Jennifer Ott, Jason Nielson, and SCIPP generally for providing the X-ray machine needed to make these measurements. This work was funded in part by the U.S. Department of Energy, Office of Science, Office of High-Energy Physics under Award Number DE-SC-0023527, and the U.S. National Science Foundation under Award Number 2146752.

\newpage
\bibliographystyle{jhep}
\bibliography{biblio.bib}






\end{document}